# Bill Kruskal and the Committee on National Statistics

**Margaret E. Martin**

In 1972 I was working at the U.S. Bureau of the Budget in the Office of Statistical Policy (formerly the Division of Statistical Standards). I had been there for some 30 years and had been appointed an assistant chief to cover the areas of employment, population and income statistics. I told the story of how I came to be executive director of the Committee on National Statistics in a 1994 interview in *Statistical Science* conducted by Miron Straf with Ingram Olkin in attendance. Let me quote from there:

> Bill Kruskal had called me to ask me if I would accept the position of Executive Director to the newly-organized Committee on National Statistics at the National Academy of Sciences, and I had turned him down. I wasn't immediately attracted to it and I didn't think I'd do a good job. So, months went by, maybe nine months, and he called me again. He was chairman of the Committee. It was in the middle of the budget season. I had just had one of these awkward incidents with Julie [Shiskin, then director of the Office of Statistical Policy] which I would have probably forgotten in two or three days, but I hadn't when Bill called me. Furthermore, I was in the middle of my 29th budget season. It was enough, you know. Bill called up and said, "I know you turned me down, but I'm calling again because I have a list of people ...." They wanted somebody who had experience with the federal government's statistical system, since few on the Committee had. He wanted to read me a list of names, and would I please comment on these people. I just heard myself saying (I hadn't thought about it at all) "I had no idea the job was still open, but if you were serious about asking me before, I'd like to be considered on the list." He said, "Well, that's it. I'm not going to read the list, then." So, I never did find out who was on the list.

Bill was a great person to work for. Looking back from this long distance I realize that he had to make great leaps between the academic views of statistics and the government administrative view. He was trying to work in two different cultures. He did a great job bridging the gap.

He contributed a great deal to the establishment of the Committee on National Statistics, and it was tough going. He hired me, but I had no experience in developing proposals or getting funds. We had one year's funding from the Russell Sage Foundation, and then one more year from them at a reduced level, but they thought the government should pay for this activity. Bill and I spent a great deal of time together going to heads of government agencies. In particular we went to the head of the National Science Foundation (NSF) and to the head of the mathematics group, where statistics was thought to be housed. They couldn't have been less interested. We finally got support from the social sciences at the NSF. Bill was always cheerful throughout this effort. Surely he was very busy at Chicago, but he called at least every week to encourage us.

He thought of himself as a one-man agency to try to draw statisticians together through his correspondence and forwarding materials to them.

When it was first established at the National Research Council (NRC), the Committee was attached to the Assembly of Mathematical and Physical Sciences. Nobody there had much experience with government statistics, but there was a section of the

*Margaret E. Martin is retired from the position of Executive Director, Committee on National Statistics, National Research Council, and lives in Mitchellville, Maryland 20721-2734, USA e-mail: mgrandaunt@aol.com.*

NRC that dealt with the life sciences—statistics was also important there—and a section that dealt with the social sciences. The staff officer in charge of social sciences was looking for additional fields to add to his group. He thought that reviewing/monitoring national policy statistics would logically fall there, either in behavioral science or in sociology or in economics. This attracted me because I had discovered that the two things that are important between the committee and its home base were (1) advice on and approval of projects and (2) suggesting appropriate people to put on individual panels for specific projects. Under the original arrangement, as staff officer I was reporting to geologists and physicists as well as mathematicians. They were kind enough to me, they didn't obstruct, but they didn't add anything because they weren't equipped to do so. So I was pleased to suggest the change, but it was against Bill's deepest feelings to leave mathematics. I admired Bill for understanding the necessity of the change. Thus it was painful for Bill to propose that we move and explain why, and to ask that nobody try to stop us, but Bill did it.

He never criticized or assessed blame, whether we fell on our faces or not. He was supportive, thoughtful, careful to try to direct us in what he thought was the right direction, but in very subtle ways. He had wide interests and wide acquaintanceship among statisticians, and these were very helpful. He laid such a sound basis for the Committee that it has flourished for over 30 years.